27th International Symposium on Superconductivity, ISS 2014

# Density-of-states fluctuation-induced negative out-of-plane magnetoresistance in overdoped Bi-2212

Tomohiro Usui[a], Shintaro Adachi[a], Takao Watanabe[a]*, Terukazu Nishizaki[b]

[a]*Graduate School of Science and Technology, Hirosaki University, Hirosaki 036-8561, Japan*
[b]*Department of Electrical Engineering and Information Technology, Kyushu Sangyo University, Fukuoka 813-8503, Japan.*

**Abstract**

We analyzed the in-plane and out-of-plane magnetoresistance (MR) for overdoped $Bi_{1.6}Pb_{0.4}Sr_2CaCu_2O_{8+\delta}$ (Bi-2212) single crystals using superconductive fluctuation theory, which considers the density-of-states (DOS) contribution in layered superconductors with the conventional s-wave pairing state. The out-of-plane results are well reproduced by the theory, implying that the large, negative out-of-plane MR as well as the sharp increase in the zero-field out-of-plane resistivity $\rho_c$ near the superconducting transition temperature $T_c$ originate from the superconductive DOS fluctuation effect. On the other hand, the in-plane results are better reproduced without the DOS contribution (i.e., using only the Aslamazov-Larkin (AL) contribution), which may be explained in terms of the d-wave superconductivity of the layered superconductors.

## 1. Introduction

The most well-known feature in the out-of-plane resistivity $\rho_c$ of high superconducting transition temperature $T_c$ (high-$T_c$) cuprates is the semiconductive upturn and the associated negative magnetoresistance (MR) upon cooling [1]. This behavior is generally thought to be caused by the opening of the pseudogap. On the other hand, a theoretical proposal [2] has suggested that this behavior is caused by the superconductive density-of-states (DOS) fluctuation effect, which dominates in the wide temperature region above $T_c$ when the system is extremely two-dimensional (2D) such as in high-$T_c$ cuprates. The quasiparticle DOS is reduced when the superconducting gap opens, resulting in a smaller out-of-plane tunneling probability and, thus, enhanced $\rho_c$. Hence, when superconductivity is suppressed by the application of magnetic fields, $\rho_c$ decreases (i.e., out-of-plane MR becomes negative). The origin of the anomalous out-of-plane transport is still in dispute.

We have recently shown, for various doping-controlled $Bi_2Sr_2CaCu_2O_{8+\delta}$ (Bi-2212), that the onset temperature of the large, negative out-of-plane MR coincides with that of the large, positive in-plane MR upon cooling [3]. Based on this observation, we have considered that the large, negative out-of-plane MR is caused by the superconductive fluctuation effect (DOS contribution) because the large, positive in-plane MR near $T_c$ may safely be attributed to the superconductive fluctuation effect (Aslamazov-Larkin (AL) contribution). Our interpretation agrees with the above theoretical proposal [2]. Although there are several reports supporting this scenario [4–7], quantitative MR studies for various samples that include the temperature or magnetic field dependences are lacking.

---

* Corresponding author. Tel.: +81-172-39-3552; fax: +81-172-39-3552.
 *E-mail address:* twatana@hirosaki-u.ac.jp

In this paper, we show the numerical analysis for both the in-plane and out-of-plane MR data of overdoped Bi-2212 using the superconductive fluctuation theory [2] including the DOS contribution effect. Then, the superconductive fluctuation contribution to the zero-field $\rho_c$ was subtracted from the experimental data. Based on the results, we discuss the origin of the semiconductive upturn for $\rho_c$.

## 2. Experimental and analysis methods

Single crystals of $Bi_{1.6}Pb_{0.4}Sr_2CaCu_2O_{8+\delta}$ were grown in air using the traveling solvent floating zone (TSFZ) method. The obtained crystals were annealed under flowing oxygen at 400°C for 50 h to promote hole doping ($T_c$ = 60 K, $p$ = 0.22). The $T_c$ was determined by the onset of zero resistivity. The doping level ($p$) was estimated using the empirical relation in ref. [8] with maximum $T_c$ = 93 K. Note that our sample is more overdoped than that of the pioneering work [4]. The in-plane resistivity $\rho_{ab}$ and $\rho_c$ measurements were performed using a DC four-terminal method. Magnetic fields **B** of up to 12 T were applied parallel to the $c$-axis.

In general, fluctuation-induced MR is composed of four different contributions: the AL contribution, DOS contribution, regular Maki-Thompson (MT(reg)) contribution, and anomalous Maki-Thompson (MT(an)) contribution. Each fluctuation type is divided into two groups: one group originates from orbital motion, and the other group originates from spin interactions (i.e., the Zeeman effect). Because the Zeeman contribution is small compared to the orbital contribution in this experimental configuration (**B** || **c**), we ignored the Zeeman contribution. Furthermore, we ignored the MT(an) contribution because it may be very small in high-$T_c$ cuprates due to strong pair-breaking effects [9] or due to d-wave superconductivity [10]. Consequently, we consider only the orbital AL, DOS, and MT(reg) contributions in this study. For the numerical calculation, we used the formulas represented in ref. [5] and ref. [11]. Note that the formulas were derived assuming the conventional s-wave superconductivity. The microscopic parameters appearing in the formulas are the Fermi velocity $v_f$ [cm/s], the quasiparticle scattering time $\tau$ [s], the mean-field transition temperature $T_{c0}$ [K], and an interlayer hopping integral $J$ [K]. The interlayer spacing $s$ was fixed to 15.4 Å throughout the analysis.

## 3. Results and discussion

Figure 1 shows the temperature dependence of $\rho_c(T)$ for $Bi_{1.6}Pb_{0.4}Sr_2CaCu_2O_{8+\delta}$ at a zero field. The slope of $\rho_c(T)$ is positive ($d\rho_c(T)/dT > 0$) for a wide temperature region, and a slight upturn occurs near $T_c$. From this result, we estimate the pseudogap opening temperature $T^*$ to be 136 K. Here, $T^*$ is defined as the temperature at which $\rho_c$ increases 1% from the high-temperature $T$-linear behavior [3]. The inset shows the expanded temperature dependence of $\rho_c(T)$ near $T_c$ for several magnetic fields up to 12 T. The negative MR rapidly increases below 90 K.

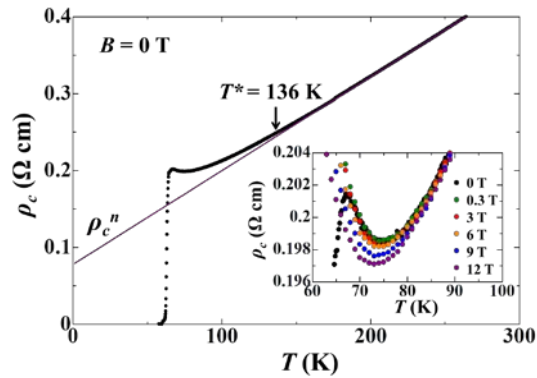

Fig. 1. Temperature dependence of the out-of-plane resistivity $\rho_c(T)$ for $Bi_{1.6}Pb_{0.4}Sr_2CaCu_2O_{8+\delta}$ ($T_c$ = 60 K). The line shows the bare resistivity $\rho_c^n$ extrapolated from the high-temperature $T$-linear behavior. The inset shows $\rho_c(T)$ in the temperature region near $T_c$ for several magnetic fields up to 12 T.

Figure 2(a) shows the temperature dependence of the out-of-plane magnetoconductivity (MC), $-\Delta\sigma_c(B, T) = -(\sigma_c(B, T) - \sigma_c(0, T))$, for $Bi_{1.6}Pb_{0.4}Sr_2CaCu_2O_{8+\delta}$ under $B$ = 12 T and the calculated results. While the AL contribution, $-\Delta\sigma_c^{AL}$, is positive and very small at higher temperatures (> 70 K), the AL contribution steeply increases below 70 K. On the other hand, the DOS contribution, $-\Delta\sigma_c^{DOS}$, is negative, and the magnitude gradually increases with decreasing temperature. The MT(reg) contribution, $-\Delta\sigma_c^{MT(reg)}$, is negligibly small. The total contribution is negative in a wide temperature region above $T_c$, which eventually becomes positive immediately near $T_c$. The experimental data is well reproduced by the calculated results, apart from the temperature interval from $T_c$ to 68 K. The discrepancy may come from sample inhomogeneity or the critical fluctuation effect [12]. Figures 2(b) and 2(c) show the experimental and calculated results of magnetic field dependence of the out-of-plane MC, $-\Delta\sigma_c(B, T)$, at 70 K and 80 K, respectively. The experimental data is well reproduced by the calculation. Figure 2(d) shows the temperature dependence of the in-plane MC, $-\Delta\sigma_{ab}(B, T) = -(\sigma_{ab}(B, T) - \sigma_{ab}(0, T))$, for $Bi_{1.6}Pb_{0.4}Sr_2CaCu_2O_{8+\delta}$ under $B = 12$ T and the calculated results. The experimental data is well reproduced by the AL contribution, $-\Delta\sigma_{ab}^{AL}$, up to $\varepsilon = 0.2$ (82 K). However, because the negative DOS and MT(reg) contributions (i.e., $-\Delta\sigma_{ab}^{DOS}$ and $-\Delta\sigma_{ab}^{MT(reg)}$) are not negligible, the total contribution apparently deviates downward from the data, especially at higher temperatures. Note that the DOS and MT(reg) contributions in this figure are shown with the opposite sign. Furthermore, even the calculated AL contribution deviates from the data above $\varepsilon = 0.2$ (82 K). This difference may be due to the normal-state MC contribution [9]. Figures 2(e) and 2(f) show the magnetic field dependence of the in-plane MC, $-\Delta\sigma_{ab}(B, T)$, at 73 K and 81 K, respectively. The experimental data is well reproduced using only the AL contribution. The reason will be discussed at the final part of this section.

Here, we assume that the pseudogap-originated MC contribution is negligible in this magnetic field range. Indeed, our tentative measurements up to 17.5 T deviated from the fitting, suggesting that the pseudogap-originated MC contribution becomes appreciable under high magnetic fields above 12 T.

We used the following parameter values for the calculation: $v_f = 2.1 \times 10^7$ cm/s, $\tau = 8.6 \times 10^{-15}$ s, $J = 20$ K, and $T_{c0} = 64$ K and 67 K for the out-of-plane and in-plane data, respectively. The $T_{c0}$ value was determined by the midpoint of each resistive transition. Using the obtained parameter values with formulas shown in ref. [13], the in-plane coherence length $\xi_{ab}$ and out-of-plane coherence length $\xi_c$ were estimated to 23 Å and 0.45 Å, respectively. This $\xi_{ab}$ value agrees with literature values [14], suggesting that the $v_f\tau$ value is reasonable. Moreover, the anisotropy $\gamma$ (= $\xi_{ab}/\xi_c$) is estimated to 51. This $\gamma$ value is smaller than $\gamma = 164$ of ref. [4]. This mainly originates because our $J$ value is larger than $J = 4.5$ K of ref. [4]. Our small $\gamma$ value is consistent with the fact that our sample is more overdoped with Pb-substitution for Bi-site. Note that the $\gamma^2$ value agrees with the observed normal-state resistivity ratio at 100 K, $\rho_c/\rho_{ab} = 2.6 \times 10^3$. From these facts, we consider that all the used parameter values are reasonable.

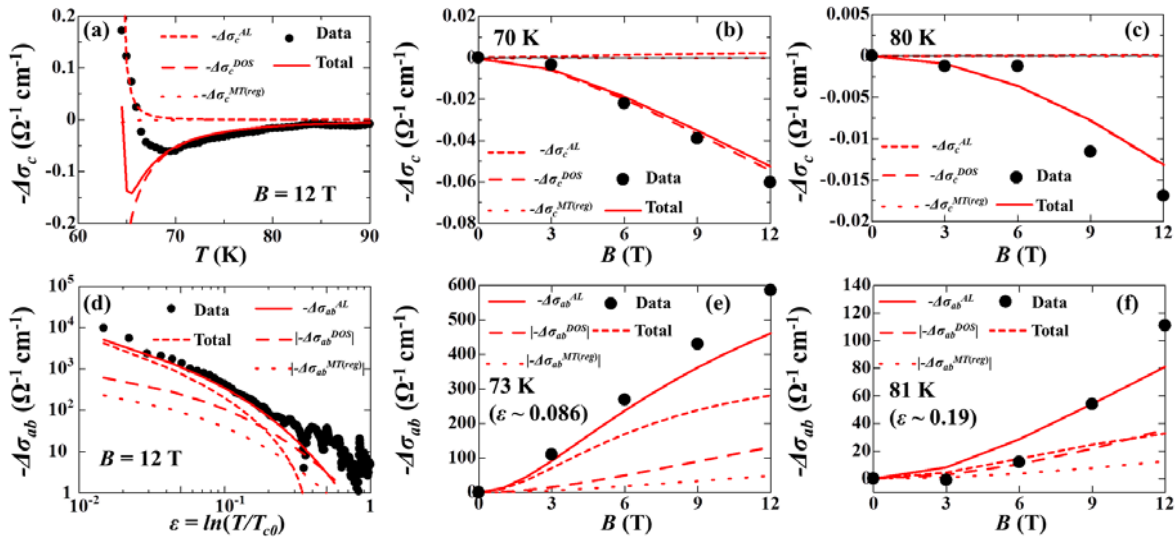

Fig. 2. (a) Temperature dependence of the out-of-plane magnetoconductivity (MC), $-\Delta\sigma_c(B, T) = -(\sigma_c(B, T) - \sigma_c(0, T))$, for $Bi_{1.6}Pb_{0.4}Sr_2CaCu_2O_{8+\delta}$. Magnetic field dependence of out-of-plane MC, $-\Delta\sigma_c(B, T)$ at (b) 70 K and (c) 80 K. The small dashed lines, large dashed lines, dotted lines, and solid lines represent the AL, DOS, MT(reg), and total contributions, respectively. (d) Temperature dependence of the in-plane magnetoconductivity (MC), $-\Delta\sigma_{ab}(B, T) = -((\sigma_{ab}(B, T) - \sigma_{ab}(0, T))$, for $Bi_{1.6}Pb_{0.4}Sr_2CaCu_2O_{8+\delta}$. Magnetic field dependence of in-plane MC, $-\Delta\sigma_{ab}(B, T)$ at (e) 73 K and (f) 81 K. The solid lines, large dashed lines, dotted lines, and small dashed lines represent the AL, DOS, MT(reg), and total contributions, respectively.

To estimate the superconductive fluctuation effect on the zero-field out-of-plane resistivity $\rho_c^{0T}$, all of the superconductive fluctuation contribution at 0 T, $\sigma_c^{fl}$, is subtracted from the experimental $\sigma_c^{0T}$ values. Figure 3 shows the subtracted result (= $1/(\sigma_c^{0T} - \sigma_c^{fl})$) as well as $\rho_c^{0T}$. The subtracted result no longer shows a steep upturn below 90 K. This result indicates that the steep upturn can be attributed to the superconductive DOS fluctuation effect, while the gradual upturn below a higher temperature is the pseudogap effect.

Finally, we discuss the reason why the negative DOS fluctuation effect could not be recognized in our in-plane data (Fig. 2(d), 2(e), 2(f)). Mori et al., reported [15] that, in the d-wave superconductors such for high-$T_c$ cuprates, an additional contribution of the positive MT(reg) form appears instead of the MT(an) contribution for $\rho_{ab}$, and the resulting MT(reg) contribution nearly cancels out the negative DOS contribution. Such features in this d-wave theory may explain our in-plane result. On the other hand, our out-of-plane result may not be altered significantly even in the d-wave approach, since the AL and MT(reg) contributions are very small compared to the DOS contribution (Fig. 2(a), 2(b), 2(c)).

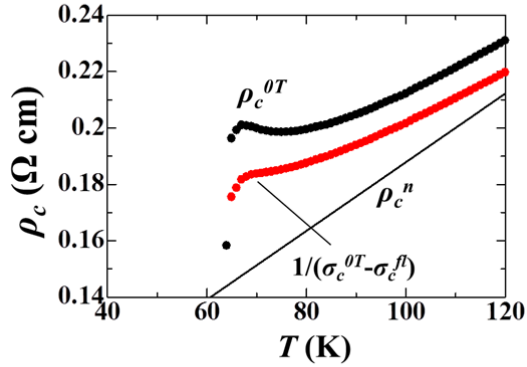

Fig. 3. The temperature dependence of the out-of-plane resistivity $\rho_c^{0T}$ at zero-field (black) and the temperature dependence of the fluctuation contribution subtracted out-of-plane resistivity $1/(\sigma_c^{0T} - \sigma_c^{fl})$ (red). Here, the normal-state bare resistivity $\rho_c^n$ is redrawn using high-temperature $T$-linear behavior of $1/(\sigma_c^{0T} - \sigma_c^{fl})$.

## 4. Conclusion

To investigate the origin of semiconductive upturn for $\rho_c$, we have analyzed the MC of overdoped Bi-2212 in the superconductive fluctuation regime using a simple model that considers the orbital AL, DOS, and MT(reg) contributions for layered s-wave superconductors. The results strongly suggest that the steep $\rho_c$ upturn and the negative MR near $T_c$ are caused by the superconductivity, in addition to the pseudogap effect at a higher temperatures. For more complete analysis, we need a d-wave superconductive fluctuation theory for the out-of-plane transport.

**Acknowledgements**

We thank N. Mori (NIT, Oyama College) for helpful discussions. The magnetoresistance measurements were performed at the High Field Laboratory for Superconducting Materials, Institute for Materials Research, Tohoku University. This work was supported by JSPS KAKENHI Grant Number 25400349.